\documentclass[prd,twocolumn,showpacs,superscriptaddress]{revtex4}
\usepackage{amsfonts}
\usepackage{amsmath}
\usepackage{amssymb}
\usepackage{bm}
\usepackage{dcolumn}
\usepackage{epsfig}
\usepackage{graphicx}
\usepackage{graphics}
\usepackage[latin1]{inputenc}
\usepackage{latexsym}
\usepackage{rotating}
\usepackage{hyperref}

\newcommand\be{\begin{equation}}
\newcommand\ba{\begin{eqnarray}}
\newcommand\ee{\end{equation}}
\newcommand\ea{\end{eqnarray}}

\begin{document}
\title[Parametrized Post-Newtonian Expansion of Chern-Simons Gravity]
{Parametrized Post-Newtonian Expansion of Chern-Simons Gravity}

\author{Stephon Alexander}
\affiliation{Center for Gravitational Wave Physics, Institute for
Gravitational Physics and Geometry and Department of Physics, 
The Pennsylvania State University, University Park, PA 16802, USA}
\author{Nicol\'as Yunes}
\affiliation{Center for Gravitational Wave Physics, Institute for
Gravitational Physics and Geometry and Department of Physics, 
The Pennsylvania State University, University Park, PA 16802, USA}

\date{\today}

\begin{abstract}
  We investigate the weak-field, post-Newtonian expansion to the
  solution of the field equations in Chern-Simons gravity with a
  perfect fluid source. In particular, we study the mapping of this
  solution to the parameterized post-Newtonian formalism to 1~PN order
  in the metric. We find that the PPN parameters of Chern-Simons
  gravity are identical to those of general relativity, with the
  exception of the inclusion of a new term that is proportional to the
  Chern-Simons coupling parameter and the curl of the PPN vector
  potentials. We also find that the new term is naturally enhanced by
  the non-linearity of spacetime and we provide a physical
  interpretation for it. By mapping this correction to the
  gravito-electro-magnetic framework, we study the corrections that
  this new term introduces to the acceleration of point particles and
  the frame-dragging effect in gyroscopic precession. We find that the
  Chern-Simons correction to these classical predictions could be used
  by current and future experiments to place bounds on intrinsic
  parameters of Chern-Simons gravity and, thus, string theory.

\end{abstract}



\maketitle

\section{Introduction}
\label{intro}

Tests of alternative theories of gravity that modify general
relativity (GR) at a fundamental level are essential to the
advancement of physics. One formalism that has had incredible success
in this task is the parameterized post-Newtonian (PPN)
framework~\cite{Schiff:1960gi,Nordtvedt:1968qs,1972ApJ...177..775N,1971ApJ...163..611W,1973ApJ...185...31W,Will:1993ns}.
In this formalism, the metric of the alternative theory is solved for
in the weak-field limit and its deviations from GR are expressed in
terms of PPN parameters. Once a metric has been obtained, one can
calculate predictions of the alternative theory, such as light
deflection and the perihelion shift of Mercury, which shall depend on
these PPN parameters. Therefore, experimental measurements of such
physical effects directly lead to constraints on the parameters of the
alternative theory. This framework, together with the relevant
experiments, have already been successfully employed to constrain
scalar-tensor theories (Brans-Dicke,
Bekenstein)~\cite{Wagoner:1970vr}, vector-tensor theories
(Will-Nordtvedt~\cite{1972ApJ...177..757W},
Hellings-Nordtvedt~\cite{PhysRevD.7.3593}), bimetric theories
(Rosen~\cite{Rosen:1974ua,1976ApJ...206..555L}) and stratified
theories (Ni~\cite{Lee:15}) (see~\cite{will:2006:cbg} for definitions
and an updated review.)

Only recently has this framework been used to study quantum
gravitational and string-theoretical inspired ideas. On the string
theoretical side, Kalyana~\cite{1994hep.th...11076K} investigated the
PPN parameters associated with the graviton-dilaton system in
low-energy string theory. More recently, Ivashchuk,
{\emph{et.~al.~}}~\cite{2001gr.qc.....1044I} studied PPN parameters in
the context of general black holes and p-brane spherically symmetric
solutions, while Bezerra, {\emph{et.~al.~}}~\cite{2002PhRvD..65j4027B}
considered domain wall spacetimes for low energy effective string
theories and derived the corresponding PPN parameters for the metric
of a wall. On the quantum gravitational side, Gleiser and
Kozameh~\cite{2001PhRvD..64h3007G} and more recently Fan,
{\emph{et.~al.~}}~\cite{Fan:2007zb} studied the possibility of testing
gravitational birefringence induced by quantum gravity, which was
proposed by Amelino-Camelia,
{\emph{el.~al.~}}~\cite{1998Natur.393..763A} and Gambini and
Pullin~\cite{Gambini:1998it}. Other non-PPN proposals have been also
put forth to test quantum gravity, for example through gravitational
waves
~\cite{will:1998:bmo,finn:2002:bmo,scharre:2002:tsg,sutton:2002:bgm,will:2004:tat,berti:2005:tgr,berti:2005:esb,Alexander:2007:bgw},
but we shall not discuss those tests here.

Chern-Simons (CS) gravity~\cite{jackiw:2003:cmo,alexander:2005:bgw} is
one such extension of GR, where the gravitational action is modified
by the addition of a parity-violating term. This extension is
promising because it is required by all $4$-dimensional
compactifications of string theory~\cite{Polchinski:1998rr} for
mathematical consistency because it cancels the Green-Schwarz
anomaly~\cite{Green:1987mn}. CS gravity, however, is not unique to
string theory and in fact has its roots in the standard model, where
it arises as a gravitational anomaly provided that there are more
flavours of left handed leptons than right handed ones.  Moreover the
CS extension to GR can arise via the embedding of the three
dimensional Chern-Simons topological current into a 4D space-time
manifold, decsribed by Jackiw and Pi \cite{jackiw:2003:cmo}

Chern-Simons gravity has been recently studied in the cosmological
context. In particular, this framework was used to shed light on the
anisotropies of the cosmic microwave background
(CMB)~\cite{Lue:1998mq,Li:2006ss,Alexander:2006mt} and the
leptogenesis
problem~\cite{alexander:2004:lfg,Alexander:2004xd,Li:2006ss}.  Parity
violation has also been shown to produce birefringent gravitational
waves~\cite{alexander:2005:bgw,Alexander:2007:bgw}, where different
polarizations modes acquire varying amplitudes. These modes obey
different propagation equations because the imaginary sector of the
classical dispersion relation is CS corrected. Different
from~\cite{Gambini:1998it}, in CS birefringence the velocity of the
gravitational wave remains that of light.

In this paper we study CS gravity in the PPN framework, extending the
analysis of~\cite{Alexander:2007zg} and providing some missing
details. In particular, we shall consider the effect of the CS
correction to the gravitational field of, for instance, a pulsar, a
binary system or a star in the weak-field limit.  These corrections
are obtained by solving the modified field equations in the weak-field
limit for post-Newtonian (PN) sources, defined as those that are
weakly-gravitating and slowly-moving~\cite{Blanchet:2002av}.  Such an
expansion requires the calculation of the Ricci and Cotton tensors to
second order in the metric perturbation. We then find that CS gravity
leads to the same gravitational field as that of classical GR and,
thus, the same PPN parameters, except for the inclusion of a new term
in the vectorial sector of the metric, namely
\be
\label{CS-correction}
g_{0i}^{(CS)} = 2 \dot{f} \left(\nabla \times V\right)_i,
\ee
where $\dot{f}$ acts as a coupling parameter of CS theory and $V_i$ is
a PPN potential. We also show that this solution can be alternatively
obtained by finding a formal solution to the modified field equations
and performing a PN expansion, as is done in PN theory. The full
solution is further shown to satisfy the additional CS constraint,
which leads to equations of motion given only by the divergence of the
stress-energy tensor.

The CS correction to the metric found here leads to an interesting
interpretation of CS gravity and forces us to consider a new type of
coupling. The interpretation consists of thinking of the field that
sources the CS correction as a fluid that permeates all of spacetime.
Then the CS correction in the metric is due to the ``dragging'' of
such a fluid by the motion of the source. Until now, couplings of the
CS correction to the angular momentum of the source had been neglected
by the string theory community.  Similarly, curl-type terms had also
been considered unnecessary in the traditional PPN framework, since
previous alternative gravity theories had not required it. As we shall
show, in CS gravity and thus in string theory, such a coupling is
naturally occurring. Therefore, a proper PPN mapping requires the
introduction of a new curl-type term with a corresponding new PPN
parameter of the type of Eq.~(\ref{CS-correction}).

A modification to the gravitational field leads naturally to
corrections of the standard predictions of GR. In order to illustrate
such a correction, we consider the CS term in the
gravito-electro-magnetic
analogy~\cite{1986bhmp.book.....T,Mashhoon:2003ax}, where we find that
the CS correction accounts for a modification of gravitomagnetism.
Furthermore, we calculate the modification to the acceleration of
point particles and the frame dragging effect in the precession of
gyroscopes. We find that these corrections are given by 
\ba
\delta a^i &=& -\frac{3}{2} \frac{\dot{f}}{r} \frac{G \, m}{c^2 \,
  r^2} \left(\frac{v}{c} \cdot n \right) \left(\frac{v}{c} \times
  n\right)^i, 
\nonumber \\
\delta\Omega^i &=& - \frac{\dot{f}}{r} \frac{G \, m}{c^3 \, r^2}
\left[ 3 \left(\frac{v}{c} \cdot n\right) n^i - \frac{v^i}{c} \right],
\ea
where $m$ and $v$ are the mass and velocities of the source, while $r$
is the distance to the source and $n^i=x^i/r$ is a unit vector, with
$\cdot$ and $\times$ the flat-space scalar and cross products.  Both
corrections are found to be naturally enhanced in regions of high
spacetime curvature. We then conclude that experiments that measure
the gravitomagnetic sector of the metric either in the weak-field
(such as Gravity Probe B~\cite{GravityProbeB}) and particularly in the
non-linear regime, will lead to a direct constraint on the CS coupling
parameter $\dot{f}$. In this paper we develop the details of how to
calculate these corrections, while the specifics of how to actually
impose a constraint, which depend on the experimental setup, are
beyond the scope of this paper.

The remainder of this paper deals with the details of the calculations
discussed in the previous paragraphs. We have divided the paper as
follows: Sec.~\ref{abc} describes the basics of the PPN framework;
Sec.~\ref{CS} discusses CS modified gravity, the modified field
equations and computes a formal solution; Sec.~\ref{PN} expands the
field equations to second order in the metric perturbation;
Sec.~\ref{solutions} iteratively solves the field equations in the PN
approximation and finds the PPN parameters of CS gravity;
Sec.~\ref{consequences} discusses the correction to the acceleration
of point particles and the frame dragging effect;
Sec.~\ref{conclusion} concludes and points to future research.

The conventions that we use throughout this work are the following:
Greek letters represent spacetime indices, while Latin letters stand
for spatial indices only; semicolons stand for covariant derivatives,
while colons stand for partial derivatives; overhead dots stand for
derivatives with respects to time. We denote uncontrolled remainders
with the symbol ${\cal{O}}(A)$, which stands for terms of order $A$.
We also use the Einstein summation convention unless otherwise
specified. Finally, we use geometrized units, where $G=c=1$, and the
metric signature $(-,+,+,+)$.

\section{The ABC of PPN}
\label{abc}

In this section we summarize the basics of the PPN framework,
following~\cite{Will:1993ns}. This framework was first developed by
Eddington, Robertson and Schiff~\cite{Will:1993ns,Schiff:1960gi}, but
it came to maturity through the seminal papers of Nordtvedt and
Will~\cite{Nordtvedt:1968qs,1972ApJ...177..775N,1971ApJ...163..611W,1973ApJ...185...31W}.
In this section, we describe the latter formulation, since it is the
most widely used in experimental tests of gravitational theories.

The goal of the PPN formalism is to allow for comparisons of different
metric theories of gravity with each other and with experiment. Such
comparisons become manageable through a slow-motion, weak-field
expansion of the metric and the equations of motion, the so-called PN
expansion. When such an expansion is carried out to sufficiently high
but finite order, the resultant solution is an accurate approximation
to the exact solution in most of the spacetime.  This approximation,
however, does break down for systems that are not slowly-moving, such
as merging binary systems, or weakly gravitating, such as near the
apparent horizons of black hole binaries. Nonetheless, as far as solar
system tests are concerned, the PN expansion is not only valid but
also highly accurate.

The PPN framework employs an order counting-scheme that is similar to
that used in multiple-scale
analysis~\cite{Bender,Kevorkian,Yunes:2005nn,Yunes:2006iw}. The symbol
${\cal{O}}(A)$ stands for terms of order $\epsilon^A$, where $\epsilon
\ll 1$ is a PN expansion parameter. For convenience, it is customary
to associate this parameter with the orbital velocity of the system
$v/c = {\cal{O}}(1)$, which embodies the slow-motion approximation. By
the Virial theorem, this velocity is related to the Newtonian
potential $U$ via $U \sim v^2$, which then implies that $U =
{\cal{O}}(2)$ and embodies the weak-gravity approximation. These
expansions can be thought of as two independent series: one in inverse
powers of the speed of light $c$ and the other in positive powers of
Newton's gravitational constant.

Other quantities, such as matter densities and derivatives, can and
should also be classified within this order-counting scheme.  Matter
density $\rho$, pressure $p$ and specific energy density $\Pi$,
however, are slightly more complicated to classify because they are
not dimensionless.  Dimensionlessness can be obtained by comparing the
pressure and the energy density to the matter density, which we assume
is the largest component of the stress-energy tensor, namely $p/\rho
\sim \Pi/\rho = {\cal{O}}(2)$. Derivatives can also be classified in
this fashion, where we find that $\partial_t/\partial_x =
{\cal{O}}(1)$. Such a relation can be derived by noting that
$\partial_t \sim v^i \nabla_i$, which comes from the Euler equations
of hydrodynamics to Newtonian order.

With such an order-counting scheme developed, it is instructive to
study the action of a single neutral particle. The Lagrangian of this
system is given by
\ba
\label{L}
L &=& \left(g_{\mu \nu} u^{\mu} u^{\nu}\right)^{1/2},
\nonumber \\
&=& \left(-g_{00} - 2 g_{0i} v^i - g_{ij} v^i v^j\right)^{1/2}
\ea
where $u^{\mu} = dx^{\mu}/dt = (1,v^i)$ is the $4$-velocity of the
particle and $v^i$ is its $3$-velocity. From Eq.~(\ref{L}), note that
knowledge of $L$ to ${\cal{O}}(A)$ implies knowledge of $g_{00}$ to
${\cal{O}}(A)$, $g_{0i}$ to ${\cal{O}}(A-1)$ and $g_{ij}$ to
${\cal{O}}(A-2)$. Therefore, since the Lagrangian is already known to
${\cal{O}}(2)$ (the Newtonian solution), the first PN correction to
the equations of motion requires $g_{00}$ to ${\cal{O}}(4)$, $g_{0i}$
to ${\cal{O}}(3)$ and $g_{ij}$ to ${\cal{O}}(2)$. Such order counting
is the reason for calculating different sectors of the metric
perturbation to different PN orders.

A PPN analysis is usually performed in a particular background, which
defines a particular coordinate system, and in an specific gauge,
called the standard PPN gauge. The background is usually taken to be
Minkowski because for solar system experiments deviations due to
cosmological effects are negligible and can, in principle, be treated
as adiabatic corrections. Moreover, one usually chooses a standard PPN
frame, whose outer regions are at rest with respect to the rest frame
of the universe. Such a frame, for example, forces the spatial sector
of the metric to be diagonal and isotropic~\cite{Will:1993ns}. The
gauge employed is very similar to the PN expansion of the Lorentz
gauge of linearized gravitational wave theory. The differences between
the standard PPN and Lorentz gauge are of ${\cal{O}}(3)$ and they
allow for the presence of certain PPN potentials in the vectorial
sector of the metric perturbation.

The last ingredient in the PPN recipe is the choice of a stress-energy
tensor. The standard choice is that of a perfect fluid, given by
\be
T^{\mu \nu} = \left( \rho + \rho \Pi + p \right) u^{\mu} u^{\nu} + p
g^{\mu \nu}.
\ee
Such a stress-energy density suffices to obtain the PN expansion of
the gravitational field outside a fluid body, like the Sun, or of
compact binary system. One can show that the internal structure of the
fluid bodies can be neglected to 1~PN order by the effacement
principle~\cite{Blanchet:2002av} in GR. Such effacement principle
might actually not hold in modified field theories, but we shall study
this subject elsewhere~\cite{Alexander:2007:wip}.

With all these machinery, on can write down a
super-metric~\cite{Will:1993ns}, namely
\ba
\label{PPN}
g_{00} &=& - 1 + 2 U - 2 \beta U^2 - 2 \xi \Phi_W + \left( 2 \gamma
  + 2 + \alpha_3 + \zeta_1 
\right. 
\nonumber \\
&-& \left. 
 2 \xi\right) \Phi_1 + 2 \left(3 \gamma - 2 \beta + 1 + \zeta_2 + \xi
\right) \Phi_2 
\nonumber \\
&+& 2 \left(1 + \zeta_3 \right) \Phi_3 + 2 \left( 3 \gamma +  3
  \zeta_4 - 2 \xi \right) \Phi_4  - \left( \zeta_1 - 2 \xi \right)
  {\cal{A}}, 
\nonumber \\
g_{0i} &=& -\frac{1}{2} \left(4 \gamma + 3 + \alpha_1 - \alpha_2 +
  \zeta_1 - 2 \xi \right) V_i 
\nonumber \\
&-& \frac{1}{2} \left(1 + \alpha_2 -  \zeta_1 + 2 \xi \right) W_i,
\nonumber\\
g_{ij} &=& \left(1 + 2 \gamma U\right) \delta_{ij},
\ea
where $\delta_{ij}$ is the Kronecker delta and where the PPN
potentials ($U,\Phi_W,\Phi_1,\Phi_2,\Phi_3,\Phi_4,{\cal{A}},V_i,W_i$)
are defined in Appendix~\ref{def-PPN-Pot}. Equation~(\ref{PPN})
describes a super-metric theory of gravity, because it reduces to
different metric theories, such as GR or other alternative
theories~\cite{Will:1993ns}, through the appropriate choice of PPN
parameters
($\gamma,\beta,\xi,\alpha_1,\alpha_2,\alpha_3,\zeta_1,\zeta_2,\zeta_3,\zeta_4$).
One could obtain a more general form of the PPN metric by performing a
post-Galilean transformation on Eq.~(\ref{PPN}), but such a procedure
shall not be necessary in this paper.

The super-metric of Eq.~(\ref{PPN}) is parameterized in terms of a
specific number of PPN potentials, where one usually employs certain
criteria to narrow the space of possible potentials to consider.  Some
of these restriction include the following: the potentials tend to
zero as an inverse power of the distance to the source; the origin of
the coordinate system is chosen to coincide with the source, such that
the metric does not contain constant terms; and the metric
perturbations $h_{00}$, $h_{0i}$ and $h_{ij}$ transform as a scalar,
vector and tensor. The above restrictions are reasonable, but, in
general, an additional subjective condition is usually imposed that is
based purely on simplicity: the metric perturbations are not generated
by gradients or curls of velocity vectors or other generalized vector
functions. As of yet, no reason had arisen for relaxing such a
condition, but as we shall see in this paper, such terms are indeed
needed for CS modified theories.

What is the physical meaning of all these parameters? One can
understand what these parameters mean by calculating the generalized
geodesic equations of motion and conservation laws~\cite{Will:1993ns}.
For example, the parameter $\gamma$ measures how much space-curvature
is produced by a unit rest mass, while the parameter $\beta$
determines how much ``non-linearity'' is there in the superposition
law of gravity.  Similarly, the parameter $\xi$ determines whether
there are preferred-location effects, while $\alpha_i$ represent
preferred-frame effects. Finally, the parameters $\zeta_i$ measure the
amount of violation of conservation of total momentum. In terms of
conservation laws, one can interpret these parameters as measuring
whether a theory is fully conservative, with linear and angular
momentum conserved ($\zeta_i$ and $\alpha_i$ vanish),
semi-conservative, with linear momentum conserved ($\zeta_i$ and
$\alpha_3$ vanish), or non-conservative, where only the energy is
conserved through lowest Newtonian order. One can verify that in GR,
$\gamma = \beta = 1$ and all other parameters vanish, which implies
that there are no preferred-location or frame effects and that the
theory is fully conservative.

A PPN analysis of an alternative theory of gravity then reduces to
mapping its solutions to Eq.~(\ref{PPN}) and then determining the PPN
parameters in terms of intrinsic parameters of the theory. The
procedure is simply as follows: expand the modified field equations in
the metric perturbation and in the PN approximation; iteratively solve
for the metric perturbation to ${\cal{O}}(4)$ in $h_{00}$, to
${\cal{O}}(3)$ in $h_{0i}$ and to ${\cal{O}}(2)$ in $h_{ij}$; compare
the solution to the PPN metric of Eq.~(\ref{PPN}) and read off the PPN
parameters of the alternative theory. We shall employ this procedure
in Sec.~\ref{solutions} to obtain the PPN parameters of CS gravity.

\section{CS Gravity in a Nutshell}
\label{CS}

In this section, we describe the basics of CS gravity, following
mainly~\cite{jackiw:2003:cmo,alexander:2005:bgw}.  In the standard CS
formalism, GR is modified by adding a new term to the gravitational
action. This term is given by~\cite{jackiw:2003:cmo}
\begin{equation}
\label{CS-action}
S_{CS} = \frac{m_{pl}^2}{64 \pi} \int d^4 x f \; \left(
{}^{\star}R \; R \right),
\end{equation}
where $m_{pl}$ is the Planck mass, $f$ is a prescribed external
quantity with units of squared mass (or squared length in geometrized
units), $R$ is the Ricci scalar and the star stands the dual operation,
such that
\begin{equation}
R{}^{\star}R = \frac{1}{2} R_{\alpha \beta \gamma \delta}
\epsilon^{\alpha \beta \mu \nu} R^{\gamma \delta}{}_{\mu \nu},
\end{equation}
with $\epsilon_{\mu \nu \delta \gamma}$ the totally-antisymmetric
Levi-Civita tensor and $R_{\mu \nu \delta \gamma}$ the Riemann tensor.

Such a correction to the gravitational action is interesting because
of the unavoidable parity violation that is introduced. Such parity
violation is inspired from CP violation in the standard model, where
such corrections act as anomaly-canceling terms. A similar scenario
occurs in string theory, where the Green-Schwarz anomaly is canceled
by precisely such a CS correction~\cite{Green:1987mn}, although CS
gravity is not exclusively tied to string theory. Parity violation in
CS gravity inexorably leads to birefringence in gravitational
propagation, where here we mean that different polarization modes obey
different propagation equations but travel at the same speed, that of
light~\cite{jackiw:2003:cmo,alexander:2004:lfg,alexander:2005:bgw,Alexander:2007:wip}.
If CS gravity were to lead to polarization modes that travel at
different speeds, then one could use recently proposed
experiments~\cite{2001PhRvD..64h3007G} to test this effect, but such
is not the case in CS gravity. Birefringent gravitational waves, and
thus CS gravity, have been proposed as possible explanations to the
cosmic-microwave-background (CMB)
anisotropies~\cite{alexander:2004:lfg}, as well as the baryogenesis
problem during the inflationary epoch~\cite{Lue:1998mq}. 

The magnitude of the CS correction is controlled by the
externally-prescribed quantity $f$, which depends on the specific
theory under consideration. When we consider CS gravity as an
effective quantum theory, then the correction is suppressed by some
mass scale $M$, which could be the electro-weak scale or some other
scale, since it is unconstrained. In the context of string theory, the
quantity $f$ has been calculated only in conservative scenarios, where
it was found to be suppressed by the Planck mass. In other scenarios,
however, enhancements have been proposed, such as in cosmologies where
the string coupling vanishes at late
times~\cite{Brandenberger:1988:sit,Tseytlin:1991:eos,Nayeri:2005:pas,sun:2006:ccm,wesley:2005:cct,Alexander:2000:bgi,Brandenberger:2001:lpi,battefeld:2006:sgc,brandenberger:2006:sgc,brandenberger:2007:sgc,brax:2004:bwc},
or where the field that generates $f$ couples to spacetime regions
with large curvature~\cite{randall:1999:lmh,randall:1999:atc} or
stress-energy density~\cite{Alexander:2007:wip,Alexander:2007:bgw}.
For simplicity, we here assume that this quantity is spatially
homogeneous and its magnitude is small but non-negligible, so that we
work to first order in the string-theoretical correction. Therefore,
we treat $\dot{f}$ as an independent perturbation
parameter,~\footnote{Formally, $\dot{f}$ by itself is dimensional, so
  it cannot be treated as an expansion parameter. A dimensionless
  parameter can, however, be constructed by dividing $\dot{f}$ by some
  length scale squared.} unrelated to $\epsilon$, the PN perturbation
parameter.

The field equations of CS modified gravity can be obtained by varying
the action with respect to the metric. Doing so, one obtains
\be
\label{EOM}
G_{\mu \nu} + C_{\mu \nu} = 8 \pi T_{\mu \nu},
\ee
where $G_{\mu \nu}$ is the Einstein tensor, $T_{\mu \nu}$ is a
stress-energy tensor and $C_{\mu \nu}$ is the Cotton tensor. The
latter tensor is defined via
\be
C_{\mu \nu} = - \frac{1}{\sqrt{-g}} \left[ f_{,\sigma}
  \epsilon^{\sigma \alpha \beta}{}_{(\mu} D_{\alpha}
R_{\nu) \beta} + \left(D_{\sigma}f_{,\tau}\right) \;
{}^{\star}R^{\tau}{}_{(\mu}{}^{\sigma}{}_{\nu)} \right],  
\ee
where parenthesis stand for symmetrization, $g$ is the determinant of
the metric, $D_{a}$ stands for covariant differentiation and colon
subscripts stand for partial differentiation. 

Formally, the introduction of such a modification to the field
equations leads to a new constraint, which is compensated by the
introduction of the new scalar field degree of freedom $f$. This
constraint originates by requiring that the divergence of the field
equations vanish, namely
\begin{equation}
\label{constraint}
D^{\mu} C_{\mu \nu} = \frac{1}{8 \sqrt{-g}} D_{\nu} f \;
\left({}^{\star}R R\right) = 0 , 
\end{equation}
where the divergence of the Einstein tensor vanished by the Bianchi
identities. If this constraint is satisfied, then the equations of
motion for the stress-energy $D_{\mu} T^{\mu \nu}$ are unaffected by
CS gravity. A common source of confusion is that
Eq.~(\ref{constraint}) is sometimes interpreted as requiring that
$R{}^{\star} R$ also vanish, which would then force the correction to
the action to vanish. However, this is not the case because, in
general, $f$ is an exact form ($d^2f=0$) and, thus,
Eq.~(\ref{constraint}) only implies an additional constraint that
forces all {\emph{solutions}} to the field equations to have a
vanishing $R{}^{\star}R$.

The previous success of CS gravity in proposing plausible explanations
to important cosmological problems prompts us to consider this
extension of GR in the weak-field regime. For this purpose, it is
convenient to rewrite the field equations in trace-reversed form,
since this form is most amenable to a PN expansion. Doing so, we find,
\be
R_{\mu \nu} + C_{\mu \nu} = 8 \pi \left(T_{\mu \nu} - \frac{1}{2}
  g_{\mu \nu} T \right),
\ee
where the trace of the Cotton tensor vanishes identically and $T =
g_{\mu \nu} T^{\mu \nu}$ is the four dimensional trace of the
stress-energy tensor. To linear order, the Ricci and Cotton tensors
are given by~\cite{jackiw:2003:cmo}
\ba
\label{linear-Lorentz-filed-eqs}
R_{\mu \nu} &=& - \frac{1}{2} \square h_{\mu \nu} + {\cal{O}}(h)^2,
\nonumber \\
C_{\mu \nu} &=& - \frac{\dot{f}}{2} \tilde{\epsilon}^{0 \alpha \beta}{}_{(\mu}
\square_{\eta} h_{\nu) \beta,\alpha} + {\cal{O}}(h)^2,
\ea
where $\tilde\epsilon^{\alpha \beta \gamma \delta}$ is the Levi-Civita
symbol, with convention $\tilde\epsilon^{0123} = + 1$, and
$\square_{\eta} = - \partial_t^2 + \eta^{ij} \partial_i \partial_j$ is
the flat space D'Alambertian, with $\eta_{\mu \nu}$ the Minkowski
metric. In Eq.~(\ref{linear-Lorentz-filed-eqs}), we have employed the
Lorentz gauge condition $h_{\mu \alpha,}{}^{\alpha} = h_{,\mu}/2$,
where $h = g^{\mu \nu} h_{\mu \nu}$ is the four dimensional trace of
the metric perturbation. 

The Cotton tensor changes the characteristic behavior of the Einstein
equations by forcing them to become third order instead of second
order. Third-order partial differential equations are common in
boundary layer theory~\cite{Bender}. However, in CS gravity, the
third-order contributions are multiplied by a factor of $f$ and we
shall treat this function as a small independent expansion parameter.
Therefore, the change in characteristics in the modified field
equations can also be treated perturbatively, which is justified
because eventhough $\dot{f}$ might be enhanced by standard model
currents, extra dimensions or a vanishing string coupling, it must
still carry some type of mass suppression.

The trace-reversed form of the field equations is useful because it
allows us to immediately find a formal solution. Inverting the
D'Alambertian operator we obtain
\be
\label{formal-sol}
{\cal{H}}_{\mu \nu} = -16 \pi \; \square_{\eta}^{-1} \left(T_{\mu \nu} -
  \frac{1}{2} g_{\mu \nu} T \right) + {\cal{O}}(h)^2,
\ee
where we have defined an effective metric perturbation as
\be
\label{effective-metric}
{\cal{H}}_{\mu \nu} \equiv h_{\mu \nu} + \dot{f} \tilde\epsilon^{0 \alpha
  \beta}{}_{(\mu} h_{\nu) \beta,\alpha}.
\ee
Note that this formal solution is identical to the formal PN solution
to the field equations in the limit $\dot{f} \to 0$. Also note that
the second term in Eq.~(\ref{effective-metric}) is in essence a curl
operator acting on the metric. This antisymmetric operator naturally
forces the trace of the CS correction to vanish, as well as the $00$
component and the symmetric spatial part. 

From the formal solution to the modified field equations, we
immediately identify the {\emph{only two possible non-zero CS
    contributions}}: a coupling to the vector component of the metric
$h_{0 i}$; and coupling to the transverse-traceless part of the
spatial metric $h_{ij}^{TT}$. The latter has already been studied in
the gravitational wave
context~\cite{jackiw:2003:cmo,alexander:2005:bgw,Alexander:2007:wip}
and it vanishes identically if we require the spatial sector of the
metric perturbation to be conformally flat. The former coupling is a
new curl-type contribution to the metric perturbation that, to our
knowledge, had so far been neglected both by the string theory and PPN
communities. In fact, as we shall see in later sections, terms of this
type will force us to introduce a new PPN parameter that is
proportional to the curl of certain PPN potentials.
  
Let us conclude this section by pushing the formal solution to the
modified field equations further to obtain a formal solution in terms
of the actual metric perturbation $h_{\mu \nu}$. Combining
Eqs.~(\ref{formal-sol}) and~(\ref{effective-metric}) we arrive at the
differential equation
\be
\label{decoup:1}
 h_{\mu \nu} + \dot{f} \tilde\epsilon^{0 \alpha
  \beta}{}_{(\mu} h_{\nu) \beta,\alpha} = -16 \pi \;
 \square_{\eta}^{-1} \left(T_{\mu \nu} - \frac{1}{2} g_{\mu \nu} T
 \right) + {\cal{O}}(h)^2.
\ee
Since we are searching for perturbations about the general
relativistic solution, we shall make the ansatz
\be
h_{\mu \nu}  = h_{\mu \nu}^{(GR)} + \dot{f} \zeta_{\mu \nu} +
{\cal{O}}(h)^2,
\ee
where $h_{\mu \nu}^{(GR)}$ is the solution predicted by general
relativity 
\be
h_{\mu \nu}^{(GR)} \equiv -16 \pi \; \square_{\eta}^{-1} \left(T_{\mu \nu} -
  \frac{1}{2} g_{\mu \nu} T \right),
\ee
and where $\zeta_{\mu\nu}$ is an unknown function we are solving for.
Inserting this ansatz into Eq.~(\ref{decoup:1}) we obtain
\be
\zeta_{\mu \nu} + \dot{f} \tilde\epsilon^{0 \alpha
  \beta}{}_{(\mu} \zeta_{\nu) \beta,\alpha} = 16 \pi \tilde\epsilon^{0
  \alpha \beta}{}_{(\mu} \partial_{\alpha} \square_{\eta}^{-1}
\left(T_{\nu) \beta} - \frac{1}{2} g_{\nu) \beta} T \right).
\ee
We shall neglect the second term on the left-hand side because it
would produce a second order correction. Such conclusion was also
reached when studying parity violation in GR to explain certain
features of the CMB~\cite{Alexander:2006mt}. We thus obtain the formal
solution
\be
\zeta_{\mu \nu} = 16 \pi \tilde\epsilon^{0
  \alpha \beta}{}_{(\mu} \partial_{\alpha} \square_{\eta}^{-1}
\left(T_{\nu) \beta} - \frac{1}{2} g_{\nu) \beta} T \right)
\ee
and the actual metric perturbation to linear order becomes
\ba
h_{\mu \nu}  &=& -16 \pi \; \square_{\eta}^{-1} \left(T_{\mu \nu} -
  \frac{1}{2} \eta_{\mu \nu} T \right) 
\\ \nonumber 
&+&
16 \pi \dot{f} \tilde\epsilon^{k \ell i}
 \square_{\eta}^{-1} \left(\delta_{i (\mu} T_{\nu) \ell,k} -
   \frac{1}{2} \delta_{i (\mu} \eta_{\nu) \ell} T_{,k} \right)  +
 {\cal{O}}(h)^2, 
\ea
where we have used some properties of the Levi-Civita symbol to
simplify this expression. The procedure presented here is general
enough that it can be directly applied to study CS gravity in the PPN
framework, as well as possibly find PN solutions to CS gravity. 

\section{PN expansion of CS Gravity}
\label{PN}

In this section, we perform a PN expansion of the field equations and
obtain a solution in the form of a PN series. This solution then
allows us to read off the PPN parameters by comparing it to the
standard PPN super-metric [Eq.~(\ref{PPN})]. In this section we shall
follow closely the methods of~\cite{Will:1993ns}
and~\cite{misner:1973:g} and indices shall be manipulated with the
Minkowski metric, unless otherwise specified.

Let us begin by expanding the field equations to second order in the
metric perturbation. Doing so we find that the Ricci and Cotton
tensors are given to second order by 
\begin{widetext}
\ba
\label{Ricci-2nd-O}
R_{\mu \nu} &=& -\frac{1}{2} \left[\square_{\eta} h_{\mu \nu} - 2
  h_{\sigma(\mu,\nu)}{}^{\sigma} +  h_{,\mu \nu} \right] 
- \frac{1}{2} \left[ h^{\rho_\lambda} \left(2 h_{\rho(\mu,\nu)\lambda}
    - h_{\mu \nu,\rho\lambda} - h_{\rho\lambda,\mu \nu} \right) 
- \frac{1}{2} h^{\rho\lambda}{}_{,\mu} h_{\rho\lambda,\nu} +
h^{\lambda}{}_{\mu,\rho} h^{\rho}{}_{\nu,\lambda}  
\right. 
\\ \nonumber 
&-& \left. 
 h^{\rho}{}_{\mu,\lambda} h_{\rho\nu,}{}^{\lambda} 
+ \frac{1}{2} \left( h^{,\lambda}  - 2 h^{\lambda \rho}{}_{,\rho} \right)
\left( h_{\mu \nu,\lambda} - 2 h_{\lambda (\mu,\nu)} \right) \right] +
{\cal{O}}(h)^3, 
\nonumber \\
C_{\mu \nu} &=& - \frac{\dot{f}}{2} \tilde{\epsilon}^{0 \alpha
  \beta}{}_{(\mu} \left( \square_{\eta} h_{\nu) \beta,\alpha} -
  h_{\sigma \beta, \alpha \nu)}{}^{\sigma} \right) - \frac{\dot{f}}{2}
\tilde{\epsilon}^{0 \alpha \beta}{}_{(\mu} \left[ h \left(\square_{\eta}
      h_{\nu) \beta,\alpha} - h_{\sigma \beta, \alpha \nu)}{}^{\sigma}
    \right) + \frac{1}{2} \left( 2 h_{\nu) (\lambda,\alpha)} -
      h_{\lambda \alpha, \nu)} 
    \right) 
\right. 
\nonumber \\
&\times& \left.
\left( \square_{\eta}h^{\lambda}{}_{\beta} - 2
      h_{\sigma}{}^{(\lambda}{}_{,\beta)}{}^{\sigma} +
      h_{,\beta}{}^{\lambda} \right) - 2 \hat{Q}
    R_{\nu) \beta,\alpha} \right]  - \frac{\dot{f}}{4}
 \tilde{\epsilon}^{\sigma \alpha \beta}{}_{(\mu} \left( 2
   h^{0}{}_{(\sigma,\tau)} - h_{\sigma \tau,}{}^{0} \right) \left(
   h^{\tau}{}_{[\beta,\alpha] \nu)} -
   h_{\nu)[\beta,\alpha]}{}^{\tau} \right)
\nonumber \\
&-& \frac{\dot{f}}{2} h_{\mu \lambda} \tilde{\epsilon}^{0 \alpha \beta
  (\lambda} \left( \square_{\eta} h_{\nu) \beta,\alpha} - h_{\sigma
    \beta, \alpha \nu)}{}^{\sigma} \right)
- \frac{\dot{f}}{2} \tilde{\epsilon}^{0 \alpha \beta
  (\mu} \left( \square_{\eta} h^{\lambda)}{}_{\beta,\alpha} -
  h_{\sigma \beta, \alpha}{}^{\sigma \lambda)} \right) h_{\nu \lambda}
  + {\cal{O}}(h)^3.
\ea
\end{widetext}
where index contraction is carried out with the Minkowski metric and
where we have not assumed any gauge condition. The operator
$\hat{Q}(\cdot)$ takes the quadratic part of its operand [of
${\cal{O}}(h)^2$] and it is explained in more detail in
Appendix~\ref{derive-Cotton}, where the derivation of the expansion of
the Cotton tensor is presented in more detail. In this derivation, we
have used the definition of the Levi-Civita tensor
\ba
\epsilon_{\alpha \beta \gamma \delta} &=& (-g)^{1/2}
\tilde\epsilon_{\alpha\beta\gamma\delta} = \left(1 - \frac{1}{2} h
  \right) \tilde\epsilon_{\alpha\beta\gamma\delta} + {\cal{O}}(h)^2,
\\ \nonumber 
\epsilon^{\alpha \beta \gamma \delta} &=& -(-g)^{-1/2}
\tilde\epsilon^{\alpha\beta\gamma\delta} = -\left(1 + \frac{1}{2} h
  \right) \tilde\epsilon^{\alpha\beta\gamma\delta} + {\cal{O}}(h)^2.
\ea
Note that the PN expanded version of the linearized Ricci tensor of
Eq.~(\ref{Ricci-2nd-O}) agrees with previous
results~\cite{Will:1993ns}. Also note that if the Lorentz condition is
enforced, several terms in both expressions vanish identically and the
Cotton tensor to first order reduces to
Eq.~(\ref{linear-Lorentz-filed-eqs}), which agrees with previous
results~\cite{jackiw:2003:cmo}.

Let us now specialize the analysis to the standard PPN gauge. For this
purpose, we shall impose the following gauge conditions
\ba
h_{jk,}{}^{k} - \frac{1}{2} h_{,j} &=& {\cal{O}}(4),
\nonumber \\
h_{0k,}{}^{k} - \frac{1}{2} h^{k}{}_{k,0} &=& {\cal{O}}(5),
\ea
where $h^{k}{}_{k}$ is the spatial trace of the metric perturbation.
Note that the first equation is the PN expansion of one of the Lorentz
gauge conditions, while the second equation is not.  This is the
reason why the previous equations where not expanded in the Lorentz
gauge.  Nonetheless, such a gauge condition does not uniquely fix the
coordinate system, since we can still perform an infinitesimal gauge
transformation that leaves the modified field equations invariant.
One can show that the Lorentz and PPN gauge are related to each other
by such a gauge transformation. In the PPN gauge, then, the Ricci
tensor takes the usual form
\ba
R_{00} &=& - \frac{1}{2} \nabla^2 h_{00} - \frac{1}{2} h_{00,i}
h_{00,}{}^{i} + \frac{1}{2} h^{ij} h_{00,ij} + {\cal{O}}(6),
\nonumber \\
R_{0i} &=& - \frac{1}{2} \nabla^2 h_{0i} - \frac{1}{4} h_{00,0i} + {\cal{O}}(5),
\nonumber \\
R_{ij} &=& - \frac{1}{2} \nabla^2 h_{ij} + {\cal{O}}(4),
\ea
which agrees with previous results~\cite{Will:1993ns}, while the
Cotton tensor reduces to
\ba
\label{expanded-Cotton}
C_{00} &=& {\cal{O}}(6),
\nonumber \\
C_{0i} &=& - \frac{1}{4} \dot{f} \tilde{\epsilon}^{0kl}{}_{i} \nabla^2
h_{0l,k} + {\cal{O}}(5), 
\nonumber \\
C_{ij} &=& - \frac{1}{2} \dot{f} \tilde{\epsilon}^{0kl}{}_{(i}
\nabla^2 h_{j) l,k} + {\cal{O}}(4),
\ea
where $\nabla = \eta^{ij} \partial_i \partial_j$ is the Laplacian of
flat space [see Appendix~\ref{derive-Cotton} for the derivation of
Eq.~(\ref{expanded-Cotton}).] Note again the explicit appearance of
two coupling terms of the Cotton tensor to the metric perturbation:
one to the transverse-traceless part of the spatial metric and the
other to the vector metric perturbation.  The PN expansions of the
linearized Ricci and Cotton tensor then allow us to solve the modified
field equations in the PPN framework.

\section{PPN Solution of CS gravity}
\label{solutions}

In this section we shall proceed to systematically solve the modified
field equation following the standard PPN iterative
procedure~\cite{Will:1993ns}.  We shall begin with the $00$ and $ij$
components of the metric to ${\cal{O}}(2)$, and then proceed with the
$0i$ components to ${\cal{O}}(3)$ and the $00$ component to
${\cal{O}}(4)$.  Once all these components have been solved for in
terms of PPN potentials, we shall be able to read off the PPN
parameters adequate to CS gravity.

\subsection{$h_{00}$ and $h_{ij}$ to ${\cal{O}}(2)$} 

Let us begin with the modified field equations for the scalar sector
of the metric perturbation. These equations are given to
${\cal{O}}(2)$ by
\be
\label{h00}
\nabla^2 h_{00} = -8 \pi \rho,  
\ee
because $T=-\rho$. Eq.~(\ref{h00}) is the Poisson equation, whose
solution in terms of PPN potentials is
\be
h_{00} = 2 U + {\cal{O}}(4). 
\ee

Let us now proceed with the solution to the field equation for the
spatial sector of the metric perturbation. This equation to
${\cal{O}}(2)$ is given by 
\be
\label{hij}
\nabla^2 h_{ij} +  \dot{f} \tilde{\epsilon}^{0kl}{}_{(i}
\nabla^2 h_{j) l,k} = -8 \pi \rho \delta_{ij},
\ee
where note that this is the first appearance of a Cotton tensor
contribution.  Since the Levi-Civita symbol is a constant and
$\dot{f}$ is only time-dependent, we can factor out the Laplacian and
rewrite this equation in terms of the effective metric
${\cal{H}}_{ij}$ as
\be
\label{effective-H}
\nabla^2 {\cal{H}}_{ij} = -8 \pi \rho \delta_{ij},
\ee
where, as defined in Sec.~\ref{CS},
\be
\label{effective-H-def}
{\cal{H}}_{ij} = h_{ij} + \dot{f} \tilde{\epsilon}^{0kl}{}_{(i} h_{j)
  l,k}.  
\ee
The solution of Eq.~(\ref{effective-H}) can be immediately found in
terms of PPN potentials as
\be
\label{effective-H-solved}
{\cal{H}}_{ij} = 2 U \delta_{ij} + {\cal{O}}(4),
\ee
which is nothing but Eq.~(\ref{formal-sol}). Recall, however, that in
Sec.~\ref{CS} we explicitly used the Lorentz gauge to simplify the
field equations, whereas here we are using the PPN gauge. The reason
why the solutions are the same is that the PPN and Lorentz gauge are
indistinguishable to this order.

Once the effective metric has been solved for, we can obtain the
actual metric perturbation following the procedure described in
Sec.~\ref{CS}. Combining Eq.~(\ref{effective-H-def}) with
Eq.~(\ref{effective-H-solved}), we arrive at the following
differential equation
\be
\label{hij-de}
h_{ij} + \dot{f} \tilde{\epsilon}^{0kl}{}_{(i} h_{j) l,k}
=  2 U \delta_{ij}. 
\ee
We look for solutions whose zeroth-order result is that predicted by
GR and the CS term is a perturbative correction, namely
\be
h_{ij} = 2 U \delta_{ij} + \dot{f} \zeta_{ij},
\ee
where $\zeta$ is assumed to be of ${\cal{O}}(\dot{f})^0$.  Inserting
this ansatz into Eq.~(\ref{hij-de}) we arrive at
\be
\zeta_{ij} +  \dot{f} \tilde{\epsilon}^{0kl}{}_{(i}
\zeta_{j) l,k} =  0,
\ee
where the contraction of the Levi-Civita symbol and the Kronecker
delta vanished. As in Sec.~\ref{CS}, note that the second term on the
left hand side is a second order correction and can thus be neglected
to discover that $\zeta_{ij}$ vanishes to this order.

The spatial metric perturbation to ${\cal{O}}(2)$ is then simply given
by the GR prediction without any CS correction, namely
\be
h_{ij} = 2 U \delta_{ij} + {\cal{O}}(4).
\ee
Physically, the reason why the spatial metric is unaffected by the CS
correction is related to the use of a perfect fluid stress-energy
tensor, which, together with the PPN gauge condition, forces the
metric to be spatially conformally flat. In fact, if the spatial
metric were not flat, then the spatial sector of the metric
perturbation would be corrected by the CS term.  Such would be the
case if we had pursued a solution to $2$ PN order, where the
Landau-Lifshitz pseudo-tensor sources a non-conformal correction to
the spatial metric~\cite{Blanchet:2002av}, or if we had searched for
gravitational wave solutions, whose stress-energy tensor
vanishes~\cite{alexander:2004:lfg,alexander:2005:bgw}. In fact, one
can check that, in such a scenario, Eq.~(\ref{effective-H}) reduces to
that found
by~\cite{jackiw:2003:cmo,alexander:2004:lfg,alexander:2005:bgw,Alexander:2007:wip}
as $\rho \to 0$. We have then found that the weak-field expansion of
the gravitational field outside a fluid body, like the Sun or a
compact binary, is unaffected by the CS correction to ${\cal{O}}(2)$.

\subsection{$h_{0i}$ to ${\cal{O}}(3)$} 

Let us now look for solutions to the field equations for the vector
sector of the metric perturbation. The field equations to
${\cal{O}}(3)$ become
\ba
\label{h0i}
\nabla^2 h_{0i} + \frac{1}{2} h_{00,0i} + \frac{1}{2} \dot{f}
\tilde\epsilon^{0kl}{}_{i} \nabla^2h_{0l,k} 
= 16 \pi \rho v_i,
\ea
where we have used that $T^{0i} = -T_{0i}$. Using the lower order
solutions and the effective metric, as in Sec.~\ref{CS}, we obtain
\ba
\label{DE-h0i}
\nabla^2 {\cal{H}}_{0i} + U_{,0i} = 16 \pi \rho v_i,
\ea
where the vectorial sector of the effective metric is
\be
\label{H0i-vector}
{\cal{H}}_{0i} =  h_{0i} + \frac{1}{2} \dot{f}
\tilde\epsilon^{0kl}{}_{i} h_{0l,k}.
\ee
We recognize Eq.~(\ref{DE-h0i}) as the standard GR field equation to
${\cal{O}}(3)$, except that the dependent function is the effective
metric instead of the metric perturbation. We can thus solve this
equation in terms of PPN potentials to obtain
\be
\label{H0i-sol}
{\cal{H}}_{0i} = -\frac{7}{2} V_i - \frac{1}{2} W_i,
\ee
where we have used that the superpotential $X$ satisfies $X_{,0j} =
V_j - W_j$ (see Appendix~\ref{def-PPN-Pot} for the definitions.)
Combining Eq.~(\ref{H0i-vector}) with Eq.~(\ref{H0i-sol}) we arrive at
a differential equation for the metric perturbation, namely
\be
 h_{0i} + \frac{1}{2} \dot{f} \tilde\epsilon^{0kl}{}_{i} h_{0l,k} =
 -\frac{7}{2} V_i - \frac{1}{2} W_i.
\ee
Once more, let us look for solutions that are perturbation about the
GR prediction, namely
\be
h_{0i} =  -\frac{7}{2} V_i - \frac{1}{2} W_i + \dot{f} \zeta_i,
\ee
where we again assume that $\zeta_i$ is of
${\cal{O}}(\dot{f})^0$. The field equation becomes 
\be
\zeta_{i} + \frac{1}{2} \dot{f} \left(\nabla \times \zeta\right)_i
= \frac{1}{2} \left( \frac{7}{2} \left(\nabla \times V\right)_i
+ \frac{1}{2} \left(\nabla \times W \right)_i \right),
\ee
where $\left(\nabla \times A\right)^i = \epsilon^{ijk} \partial_j A_k$
is the standard curl operator of flat space. As in Sec.~\ref{CS}, note
once more that the second term on the left-hand side is again a second
order correction and we shall thus neglect it. Also note that the curl
of the $V_i$ potential happens to be equal to the curl of the $W_i$
potential. The solution for the vectorial sector of the actual
gravitational field then simplifies to
\be
h_{0i} =  -\frac{7}{2} V_i - \frac{1}{2} W_i + 2 \dot{f} \left(\nabla
  \times V\right)_i + {\cal{O}}(5).
\ee

We have arrived at the first contribution of CS modified gravity to
the metric for a perfect fluid source.  Chern-Simons gravity was
previously seen to couple to the transverse-traceless sector of the
metric perturbation for gravitational wave
solutions~\cite{jackiw:2003:cmo,alexander:2004:lfg,alexander:2005:bgw,Alexander:2007:wip}.
The CS correction is also believed to couple to Noether vector
currents, such as neutron currents, which partially fueled the idea
that this correction could be enhanced. However, to our knowledge,
this correction was never thought to couple to vector metric
perturbations. From the analysis presented here, we see that in fact
CS gravity does couple to such terms, even if the matter source is
neutrally charged. The only requirement for such couplings is that the
source is not static, {\emph{ie.}} that the object is either moving or
spinning relative to the PPN rest frame so that the PPN vector
potential does not vanish.  The latter is suppressed by a relative
${\cal{O}}(1)$ because in the far field the velocity of a compact
object produces a term of ${\cal{O}}(3)$ in $V_i$, while the spin
produces a term of ${\cal{O}}(4)$. In a later section, we shall
discuss some of the physical and observational implications of such a
modification to the metric.

\subsection{$h_{00}$ to ${\cal{O}}(4)$} 
A full analysis of the PPN structure of a modified theory of gravity
requires that we solve for the $00$ component of the metric
perturbation to ${\cal{O}}(4)$. The field equations to this order are 
\ba
\label{h00:2}
&-& \frac{1}{2} \nabla^2 h_{00} - \frac{1}{2} h_{00,i} h_{00,i} +
\frac{1}{2} h_{ij} h_{00,ij} = 4 \pi \rho \left[1 
\right. 
\nonumber \\
&+& \left. 
 2 \left( v^2 - U + \frac{1}{2} \Pi + \frac{3}{2} \frac{p}{\rho}
 \right)\right],
\ea
where the CS correction does not contribute at this order (see
Appendix~\ref{derive-Cotton}.)  Note that the $h_{0i}$ sector of the
metric perturbation to ${\cal{O}}(3)$ does not feed back into the
field equations at this order either. The terms that do come into play
are the $h_{00}$ and $h_{ij}$ sectors of the metric, which are not
modified to lowest order by the CS correction. The field equation,
thus, reduce to the standard one of GR, whose solution in terms of PPN
potentials is
\be
h_{00} = 2 U - 2 U^2 + 4 \Phi_1 + 4 \Phi_2 + 2 \Phi_3 + 6 \Phi_4 +
{\cal{O}}(6). 
\ee
We have thus solved for all components of the metric perturbation to 1
PN order beyond the Newtonian answer, namely $g_{00}$ to
${\cal{O}}(4)$, $g_{0i}$ to ${\cal{O}}(3)$ and $g_{ij}$ to
${\cal{O}}(2)$.

\subsection{PPN Parameters for CS  Gravity}
We now have all the necessary ingredients to read off the PPN
parameters of CS modified gravity. Let us begin by writing the full
metric with the solutions found in the previous subsections:
\ba
\label{CS-full-metric}
g_{00} &=& -1 + 2 U - 2 U^2 + 4 \Phi_1 + 4 \Phi_2 + 2 \Phi_3 + 6 \Phi_4 +
{\cal{O}}(6), 
\nonumber \\
g_{0i} &=& -\frac{7}{2} V_i - \frac{1}{2} W_i + 2 \dot{f} \left(\nabla
  \times V\right)_i + {\cal{O}}(5),
\nonumber \\
g_{ij} &=& \left(1 + 2 U \right) \delta_{ij} + {\cal{O}}(4). 
\ea
One can verify that this metric is indeed a solution of
Eqs.~(\ref{h00}), (\ref{hij}), (\ref{h0i}) and~(\ref{h00:2}) to the
appropriate PN order and to first order in the CS coupling parameter.
Also note that the solution of Eq.~(\ref{CS-full-metric})
automatically satisfies the constraint $^{\star}R R = 0$ to linear
order because the contraction of the Levi-Civita symbol with two
partial derivatives vanishes. Such a solution is then allowed in CS
gravity, just as other classical solutions are~\cite{Guarrera:2007tu},
and the equations of motion for the fluid can be obtained directly
from the covariant derivative of the stress-energy tensor.

We can now read off the PPN parameters of the CS modified theory by
comparing Eq.~(\ref{PPN}) to Eq.~(\ref{CS-full-metric}). A visual
inspection reveals that the CS solution is identical to the classical
GR one, which implies that $\gamma = \beta = 1$, $\zeta = 0$ and
$\alpha_1 = \alpha_2 = \alpha_3 = \xi_1 = \xi_2 = \xi_3 = \xi_4 = 0$
and there are no preferred frame effects. However, Eq.~(\ref{PPN})
contains an extra term that cannot be modeled by the standard PPN
metric of Eq.~(\ref{PPN}), namely the curl contribution to $g_{0i}$.
We then see that the PPN metric must be enhanced by the addition of a
curl-type term to the $0i$ components of the metric, namely
\ba
g_{0i} &\equiv& -\frac{1}{2} \left(4 \gamma + 3 + \alpha_1 - \alpha_2 +
  \zeta_1 - 2 \xi \right) V_i 
\nonumber \\
&-& \frac{1}{2} \left(1 + \alpha_2 -  \zeta_1 + 2 \xi \right) W_i +
\chi \left(r \nabla \times V\right)_{i},
\ea
where $\chi$ is a {\emph{new}} PPN parameter and where we have
multiplied the curl operator by the radial distance to the source,
$r$, in order to make $\chi$ a proper dimensionless parameter. Note
that there is no need to introduce any additional PPN parameters
because the curl of $W_i$ equals the curl of $V_i$. In fact, we could
have equally parameterized the new contribution to the PPN metric in
terms of the curl of $W_i$, but we chose not to because $V_i$ appears
more frequently in PN theory. For the case of CS modified gravity,
the new $\chi$ parameter is simply
\be
\chi = 2 \frac{\dot{f}}{r}, 
\ee
which is dimensionless since $\dot{f}$ has units of length. If an
experiment could measure or place bounds on the value of $\chi$, then
$\dot{f}$ could also be bounded, thus placing a constraint on the CS
coupling parameter.

\section{Astrophysical Implications}
\label{consequences}

In this section we shall propose a physical interpretation to the CS
modification to the PPN metric and we shall calculate some GR
predictions that are modified by this correction. This section,
however, is by no means a complete study of all the possible
consequences of the CS correction, which is beyond the scope of this
paper.

Let us begin by considering a system of $A$ nearly spherical bodies,
for which the gravitational vector potentials are
simply~\cite{Will:1993ns}
\ba
\label{vector-pot}
V^i &=& \sum_A \frac{m_A}{r_A} v^i_A + \frac{1}{2} \sum_A
\left(\frac{J_A}{r_A^2} \times n_A\right)^i,
\\ \nonumber 
W^i &=& \sum_A \frac{m_A}{r_A} \left(v_A \cdot n_A\right) n_A^i +
\frac{1}{2} \sum_A \left(\frac{J_A}{r_A^2} \times n_A\right)^i,
\ea
where $m_A$ is the mass of the $A$th body, $r_A$ is the field point
distance to the $A$th body, $n_A^i = x^i_A/r_A$ is a unit vector
pointing to the $A$th body, $v_A$ is the velocity of the $A$th body
and $J_A^i$ is the spin-angular momentum of the $A$th body. For
example, the spin angular momentum for a Kerr spacetime is given by
$J^i = m^2 a^i$, where $a$ is the dimensionless Kerr spin parameter.
Note that if $A = 2$ then the system being modeled could be a binary
of spinning compact objects, while if $A=1$ it could represent the
field of the sun or that of a rapidly spinning neutron star or pulsar.

In obtaining Eq.~(\ref{vector-pot}), we have implicitly assumed a
point-particle approximation, which in classical GR is justified by
the effacement principle. This principle postulates that the internal
structure of bodies contributes to the solution of the field equations
to higher PN order. One can verify that this is indeed the case in
classical GR, where internal structure contributions appear at $5$ PN
order. In CS gravity, however, it is a priori unclear whether an
analogous effacement principle holds because the CS term is expected
to couple with matter current via standard model-like interactions. If
such is the case, it is possible that a ``mountain'' on the surface of
a neutron star~\cite{Owen:2005fn} or an r-mode
instability~\cite{Chandrasekhar:1992pr,Friedman:1978hf,Lindblom:1998wf}
enhances the CS contribution. In this paper, however, we shall neglect
these interactions, and relegate such possibilities to future
work~\cite{Alexander:2007:wip}.

With such a vector potential, we can calculate the CS correction to
the metric. For this purpose, we define the correction $\delta g_{0i}
\equiv g_{0i} - g_{0i}^{(GR)}$, where $g_{0i}^{(GR)}$ is the GR
prediction without CS gravity. We then find that the CS corrections is
given by 
\be
\label{CS-term}
\delta g_{0i} = 2  \sum_A \frac{\dot{f}}{r_A} \left[ \frac{m_A}{r_A}
  \left(v_A \times n_A \right)^i - \frac{J^i_A}{2 r_A^2} + \frac{3}{2}
  \frac{\left(J_A \cdot n_A\right)}{r_A^2} n_A^i \right],
\ee
where the $\cdot$ operator is the flat-space inner product and where
we have used the identities $\tilde\epsilon_{ijk} \tilde\epsilon_{klm}
= \delta_{il} \delta_{jm} - \delta_{im} \delta_{jl}$ and
$\tilde\epsilon_{ilk} \tilde\epsilon_{jlm} = 2 \delta_{ij}$.  Note
that the first term of Eq.~(\ref{CS-term}) is of ${\cal{O}}(3)$, while
the second and third terms are of ${\cal{O}}(4)$ as previously
anticipated. Also note that $\dot{f}$ couples both to the spin and
orbital angular momentum. Therefore, whether the system under
consideration is the Solar system ($v^i$ of the Sun is zero while
$J^i$ is small), the Crab pulsar ($v^i$ is again zero but $J^i$ is
large) or a binary system of compact objects (neither $v^i$ nor $J^i$
vanish), there will in general be a non-vanishing coupling between the
CS correction and the vector potential of the system.

From the above analysis, it is also clear that the CS correction
increases with the non-linearity of the spacetime. In other words, the
CS term is larger not only for systems with large velocities and
spins, but also in regions near the source. For a binary system, this
fact implies that the CS correction is naturally enhanced in the last
stages of inspiral and during merger. Note that this enhancement is
{\emph{different}} from all previous enhancements proposed, since it
does not require the presence of
charge~\cite{Alexander:2007:wip,Alexander:2007:bgw}, a fifth dimension
with warped
compactifications~\cite{randall:1999:lmh,randall:1999:atc}, or a
vanishing string
coupling~\cite{Brandenberger:1988:sit,Tseytlin:1991:eos,Nayeri:2005:pas,sun:2006:ccm,wesley:2005:cct,Alexander:2000:bgi,Brandenberger:2001:lpi,battefeld:2006:sgc,brandenberger:2006:sgc,brandenberger:2007:sgc,brax:2004:bwc}.
Unfortunately, the end of the inspiral stage coincides with the edge
of the PN region of validity and, thus, a complete analysis of such a
natural enhancement will have to be carried out through numerical
simulations.

In the presence of a source with the vector potentials of
Eq.~(\ref{vector-pot}), we can write the vectorial sector of the
metric perturbation in a suggestive way, namely
\ba
\label{effective}
g_{0i} &=& \sum_A - \frac{7}{2} \frac{m_A}{r_A} v^i_A 
- \sum_A \frac{m_A}{6 r_A^2} \left(v_A - v_A^{(eff)}\right)^i
\\ \nonumber 
&-& \frac{1}{2} \sum_A n_A^i \frac{m_A}{r_A} v^{(eff)}_A \cdot n_A  
- 2 \sum_A \left[ \frac{J_A^{(eff)}}{r_A^2} \times n_A \right]^i,
\ea
where we have defined an effective velocity and angular momentum
vector via
\ba
v_{A(eff)}^i &=& v^i_A - 6 \dot{f} \frac{J^i_A}{m_A r_A^2},
\nonumber \\
J_{A(eff)}^i &=& J^i_A - \dot{f} m_A v_A^i,
\ea
or in terms of the Newtonian orbital angular momentum $L^i_{A (N)} =
r_A \times p_A$ and linear momentum $p^i_{A (N)} = m_A v_A^i$
\ba
L^i_{A(eff)} &=& L^i_{A(N)} - 6 \dot{f} \left(n_A \times J_A\right)^i,
\nonumber \\
J^i_{A(eff)} &=& J_A^i - \dot{f} p_A^i.
\ea
From this analysis, it is clear that the CS corrections seems to
couple to both a quantity that resembles the orbital and the spin
angular momentum vector. Note that when the spin angular momentum
vanishes the vectorial metric perturbation is identical to that of a
spinning moving fluid, but where the spin is induced by the coupling
of the orbital angular momentum to the CS term. 

The presence of an effective CS spin angular momentum in non-spinning
sources leads to an interesting physical interpretation.  Let us model
the field that sources $\dot{f}$ as a fluid that permeates all of
spacetime. This field could be, for example, a model-independent
axion, inspired by the quantity introduced in the standard model to
resolve the strong CP problem~\cite{Dine:1981rt}. In this scenario,
then the fluid is naturally ``dragged'' by the motion of any source
and the CS modification to the metric is nothing but such dragging.
This analogy is inspired by the ergosphere of the Kerr solution, where
inertial frames are dragged with the rotation of the black hole. In
fact, one could push this analogy further and try to construct the
shear and bulk viscosity of such a fluid, but we shall not attempt
this here. Of course, this interpretation is to be understood only
qualitatively, since its purpose is only to allow the reader to
picture the CS modification to the metric in physical terms.

An alternative interpretation can be given to the CS modification in
terms of the gravito-electro-magnetic (GED)
analogy~\cite{1986bhmp.book.....T,Mashhoon:2003ax}, which shall allow
us to easily construct the predictions of the modified theory. In this
analogy, one realizes that the PN solution to the linearized field
equations can be written in terms of a potential and vector potential,
namely
\be
ds^2 = - \left(1 - 2 \Phi \right) dt^2 - 4 \left(A \cdot dx \right) dt +
\left(1 + 2 \Phi \right) \delta_{ij} dx^i dx^j,
\ee
where $\Phi$ reduces to the Newtonian potential $U$ in the Newtonian
limit~\cite{Mashhoon:2003ax} and $A^i$ is a vector potential related
to the metric via $A_i = -g_{0i}/4$. One can then construct GED fields
in analogy to Maxwell's electromagnetic theory via
\ba
E^i &=& - \left(\nabla\Phi\right)^i - \partial_t \left(\frac{1}{2} A^i \right), 
\nonumber \\
B^i &=& \left(\nabla \times A\right)^i,
\ea
which in terms of the vectorial sector of the metric perturbation
becomes
\ba
E^i &=& - \left(\nabla\Phi\right)^i + \frac{1}{8} \dot{g}^i,
\nonumber \\
B^i &=& -\frac{1}{4} \left(\nabla \times g\right)^i,
\ea
where we have defined the vector $g^i = g_{0i}$. The geodesic
equations for a test particle then reduce to the Lorentz force law,
namely
\ba
F^i = - m E^i  - 2 m \left(v \times B\right)^i.
\ea

We can now work out the effect of the CS correction on the GED fields
and equations of motion. First note that the CS correction only
affects $g$. We can then write the CS modification to the Lorentz
force law by defining $\delta a^i = a^i - a^i_{(GR)}$, where
$a^i_{(GR)}$ is the acceleration vector predicted by GR, to obtain,
\be
\delta a^i = \frac{1}{8} \delta\dot{g}^i + \frac{1}{2} \left(v
  \times \delta\Omega\right)^i,
\ee
where we have defined the angular velocity
\be
\delta \Omega^i = \left(\nabla \times \delta g\right)^i.
\ee
The time derivative of the vector $g^i$ is of ${\cal{O}}(5)$ and can
thus be neglected, but the angular velocity cannot and it is given by
\be
\label{ang-vel}
\delta\Omega^i = - \sum_A \dot{f} \frac{m_A}{r_A^3} \left[ 3 \left(v_A \cdot
    n_A\right) n_A^i - v_A^i \right],
\ee
which is clearly of ${\cal{O}}(3)$. Note that although the first term
between square brackets cancels for circular orbits because $n_A^i$ is
perpendicular to $v_A^i$ to Newtonian order, the second term does not.
The angular velocity adds a correction to the acceleration of
${\cal{O}}(4)$, namely
\be
\label{acc}
\delta a^i = -\frac{3}{2} \sum_A \dot{f} \frac{m_A}{r_A^3} \left(v_A
  \cdot n_A \right) \left(v_A \times n_A\right)^i,
\ee
which for a system in circular orbit vanishes to Newtonian order. One
could use this formalism to find the perturbations in the motion of
moving objects by integrating Eq.~(\ref{acc}) twice.  However, for
systems in a circular orbit, such as the Earth-Moon system or compact
binaries, this correction vanishes to leading order. Therefore, lunar
ranging experiments~\cite{Murphy:2007nt} might not be able to
constraint $\dot{f}$.

Another correction to the predictions of GR is that of the precession
of gyroscopes by the so-called Lense-Thirring or frame-dragging
effect. In this process, the spin angular momentum of a source twists
spacetime in such a way that gyroscopes are dragged with it. The
precession angular velocity depends on the vector sector of the metric
perturbation via Eq.~(\ref{ang-vel}). Thus, the full Lense-Thirring
term in the precession angular velocity of precessing gyroscopes is
\be 
\Omega^i_{LT} = - \frac{1}{r_A^3} \sum_A J_{A(eff)}^i - 3 n^i_A
\left(J_{A(eff)} \cdot n_A \right)^i.  
\ee
Note that this angular velocity is identical to the GR prediction,
except for the replacement $J_A^i \to J^i_{A(eff)}$. In CS modified
gravity, then, the Lense-Thirring effect is not only produced by the
spin angular momentum of the gyroscope but also by the orbital angular
momentum that couples to the CS correction. Therefore, if an
experiment were to measure the precession of gyroscopes by the
curvature of spacetime (see, for example, Gravity Probe
B~\cite{GravityProbeB}) one could constraint $\dot{f}$ and thus some
intrinsic parameters of string theory. Note, however, that the CS
correction depends on the velocity of the bodies with respect to the
inertial PPN rest-frame. In order to relate these predictions to the
quantities that are actually measured in the experiment, one would
have to transform to the experiment's frame, or perhaps to a basis
aligned with the direction of distant stars~\cite{Will:1993ns}.

Are there other experiments that could be performed to measure such a
deviation from GR? Any experiment that samples the vectorial sector of
the metric would in effect be measuring such a deviation.  In this
paper, we have only discussed modifications to the frame-dragging
effect and the acceleration of bodies through the GED analogy, but
this need not be the only corrections to classical GR predictions. In
fact, any predictions that depends on $g_{0i}$ indirectly, for example
via Christoffel symbols, will probably also be modified unless the
corrections is fortuitously canceled. In this paper, we have laid the
theoretical foundations of the weak-field correction to the metric due
to CS gravity and studied some possible corrections to classical
predictions. A detailed study of other corrections is beyond the scope
of this paper.

\section{Conclusion}
\label{conclusion}
We have studied the weak-field expansion of the solution to the CS
modified field equations in the presence of a perfect fluid PN source
in the point particle limit.  Such an expansion required that we
linearize the Ricci and Cotton tensor to second order in the metric
perturbation without any gauge assumption. An iterative PPN formalism
was then employed to solve for the metric perturbation in this
modified theory of gravity.  We have found that CS gravity possesses
the same PPN parameters as those of GR, but it also requires the
introduction of a new term and PPN parameter that is proportional to
the curl of the PPN vector potentials. Such a term is enhanced in
non-linear scenarios without requiring the presence of standard model
currents, large extra dimensions or a vanishing string coupling.

We have proposed an interpretation for the new term in the metric
produced by CS gravity and studied some of the possible consequences
it might have on GR predictions. The interpretation consists of
picturing the field that sources the CS term as a fluid that permeates
all of spacetime. In this scenario, the CS term is nothing but the
``dragging'' of the fluid by the motion of the source.  Irrespective
of the validity of such an interpretation, the inclusion of a new term
to the weak-field expansion of the metric naturally leads to
corrections to the standard GR predictions.  We have studied the
acceleration of point particles and the Lense-Thirring contribution to
the precession of gyroscopes. We have found that both corrections are
proportional to the CS coupling parameter and, therefore, experimental
measurement of these effects might be used to constraint CS and,
possibly, string theory.

Future work could concentrate on studying further the non-linear
enhancement of the CS correction and the modifications to the
predictions of GR. The PPN analysis performed here breaks down very
close to the source due to the use of a point particle approximation
in the stress energy tensor. One possible research route could consists
of studying the CS correction in a perturbed Kerr
background~\cite{Yunes:2005ve}. Another possible route could be to
analyze other predictions of the theory, such as the perihelion shift
of Mercury or the Nordtvedt effect. Furthermore, in light of the
imminent highly-accurate measurement of the Lense-Thirring effect by
Gravity Probe B, it might be useful to revisit this correction in a
frame better-adapted to the experimental setup.  Finally, the CS
modification to the weak-field metric might lead to non-conservative
effects and the breaking of the effacement
principle~\cite{Alexander:2007:wip}, which could be studied through
the evaluation of the gravitational pseudo stress-energy tensor.
Ultimately, it will be experiments that will determine the viability
of CS modified gravity and string theory.

\acknowledgments The authors acknowledge the support of the Center for
Gravitational Wave Physics funded by the National Science Foundation
under Cooperative Agreement PHY-01-14375, and support from NSF grants
PHY-05-55-628. We would also like to thank Cliff Will for encouraging
one of us to study the PPN formalism and Pablo Laguna for suggesting
one of us to look into the PPN expansion of CS gravity. We would also
like to thank R.~Jackiw, R.~ Wagoner and Ben Owen for enlightening
discussions and comments.

\appendix

\section{PPN Potentials}
\label{def-PPN-Pot}
In this appendix, we present explicit expressions for the PPN
potentials used to parameterize the metric in Eq.~(\ref{PPN}). These
potentials are the following:
\ba
U &\equiv& \int \frac{\rho}{|{\bf{x}} - {\bf{x}}'|}
d^3 x',
\nonumber \\
V_i &\equiv& \int \frac{\rho' v_i'}{|{\bf{x}} -
  {\bf{x}}'|} d^3 x',
\nonumber \\
W_i &\equiv& \int \frac{\rho' v_j'(x - x')^j
(x - x')_i}{|{\bf{x}} -  {\bf{x}}'|^3} d^3 x',
\nonumber \\
\Phi_W &\equiv& \int \rho' \rho'' \frac{(x - x')^i}{|{\bf{x}} -
  {\bf{x}}'|^3} \left( \frac{(x' - x'')_i}{|{\bf{x}} -
  {\bf{x}}''|} - \frac{(x - x'')_i}{|{\bf{x}}' -
  {\bf{x}}''|} \right) d^3 x' d^3 x'',
\nonumber \\
\Phi_1 &\equiv& \int \frac{\rho' v'^2}{|{\bf{x}} - {\bf{x}}'|} d^3 x',
\qquad 
\Phi_2 \equiv \int \frac{\rho' U'}{|{\bf{x}} - {\bf{x}}'|} d^3 x',
\nonumber \\
\Phi_3 &\equiv& \int \frac{\rho' \Pi'}{|{\bf{x}} - {\bf{x}}'|} d^3 x',
\qquad 
\Phi_4 \equiv \int \frac{p'}{|{\bf{x}} - {\bf{x}}'|} d^3 x',
\nonumber \\
{\cal{A}} &\equiv& \int \frac{\rho' \left[v'_i \left(x -
      x'\right)^i\right]^2}{|{\bf{x}} - {\bf{x}}'|} d^3 x', 
\nonumber \\
X &\equiv& \int \rho' |{\bf{x}} - {\bf{x}}'| d^3 x'.
\ea
These potentials satisfy the following relations 
\ba
\nabla^2 U &=& -4 \pi \rho,
\qquad 
\nabla^2 V_i = -4 \pi \rho v_i,
\nonumber \\
\nabla^2 \Phi_1 &=& -4 \pi \rho v^2,
\qquad 
\nabla^2 \Phi_2 = -4 \pi \rho U,
\nonumber \\
\nabla^2 \Phi_3 &=& -4 \pi \rho \Pi,
\qquad 
\nabla^2 \Phi_4 = -4 \pi p,
\nonumber \\
\nabla^2  X &=& -2 U
\ea
The potential $X$ is sometimes referred to as the super-potential
because it acts as a potential for the Newtonian potential.

\section{Linearization of the Cotton Tensor}
\label{derive-Cotton}

In this appendix, we present some more details on the derivation of
the linearized Cotton tensor to second order. We begin with the
definition of the Cotton tensor~\cite{jackiw:2003:cmo} in terms of the
symmetrization operator, namely
\be
C^{\mu \nu} = - \frac{1}{\sqrt{-g}} \left[ \left(D_{\sigma} f\right)
  \epsilon^{\sigma \alpha \beta (\mu} D_{\alpha} R^{\nu)}{}_{\beta} +
\left(D_{\sigma \tau}f\right) ^{\star}R^{\tau (\mu| \sigma |\nu)} \right].
\ee
Using the symmetries of the Levi-Civita and Riemann tensor, as well as
the fact that $f$ depends only on time, we can simplify the Cotton
tensor to
\ba
C^{\mu \nu} &=& (-g)^{-1} \dot{f} \left[ \tilde\epsilon^{0 \alpha
    \beta (\mu} R^{\nu)}{}_{\beta,\alpha} + \tilde\epsilon^{0 \alpha
    \beta (\mu} \Gamma^{\nu)}_{\lambda \alpha} R^{\lambda}_{\beta}
\right. 
\nonumber \\
&+& \left. 
 \frac{1}{2} \Gamma^{0}_{\sigma \tau} \tilde\epsilon^{\sigma \alpha
  \beta (\mu} R^{\nu) \tau}{}_{\alpha \beta}\right].
\ea
Noting that the determinant of the metric is simply $g = -1 + h$, so
that $(-g)^{-1} = 1 + h$, we can identify four terms in the Cotton
tensor
\ba
C^{\mu \nu}_A &=& \dot{f} \tilde\epsilon^{0 \alpha
    \beta (\mu} \left[\hat{L}R^{\nu)}{}_{\beta,\alpha}\right],
\nonumber \\
C^{\mu \nu}_B &=& \dot{f} \tilde\epsilon^{0 \alpha \beta (\mu}
h_{\rho \rho} \left[\hat{L}R^{\nu)}{}_{\beta,\alpha}\right],
\nonumber \\
C^{\mu \nu}_C &=& \dot{f} \tilde\epsilon^{0 \alpha
    \beta (\mu} \left[\hat{L}\Gamma^{\nu)}_{\lambda \alpha}\right]
  \left[\hat{L}R^{\lambda}{}_{\beta}\right], 
\nonumber \\
C^{\mu \nu}_D &=& \frac{\dot{f}}{2} \tilde\epsilon^{\sigma \alpha
  \beta (\mu} \left[\hat{L}\Gamma^{0}_{\sigma \tau}\right]
\left[\hat{L}R^{\nu) \tau}{}_{\alpha \beta}\right],
\nonumber \\
C^{\mu \nu}_E &=& \dot{f} \tilde\epsilon^{0 \alpha
    \beta (\mu} \left[\hat{Q}R^{\nu)}{}_{\beta,\alpha}\right],
\ea
where the $\hat{L}$ operator stands for the linear part of its
operand, while the $\hat{Q}$ operator isolates the quadratic part of
its operand. For example, if we act $\hat{L}$ and $\hat{Q}$ on $(1 +
h)^n$, where $n$ is some integer, we obtain
\ba
\left[\hat{L} (1 + h)^n\right] &=& n h,
\;
\left[\hat{Q} (1 + h)^n\right] = \frac{n(n-1)}{2} h^2.
\ea

Let us now compute each of these terms separately. The first four
terms are given by 
\ba
C^{\mu \nu}_A &=& - \frac{\dot{f}}{2} \tilde{\epsilon}^{0 \alpha \beta
  (\mu} \left( \square_{\eta} h^{\nu)}{}_{\beta,\alpha} - h_{\sigma
    \beta,}{}^{\nu}{}_{\alpha \sigma} \right),
\nonumber \\
C^{\mu \nu}_B &=& - \frac{\dot{f}}{2} h \tilde{\epsilon}^{0 \alpha \beta
  (\mu} \left( \square_{\eta} h^{\nu)}{}_{\beta,\alpha} - h_{\sigma
    \beta,}{}^{\nu}{}_{\alpha \sigma} \right),
\nonumber \\
C^{\mu \nu}_C &=& - \frac{\dot{f}}{4} \tilde{\epsilon}^{0 \alpha \beta
  (\mu} \left( h^{\nu)}{}_{\lambda,\alpha} +
  h^{\nu}{}_{\alpha,\lambda} - h_{\lambda \alpha,}{}^{\nu)} \right)
\nonumber \\
&\times&
\left(\square_{\eta}h^{\lambda}{}_{\beta} -
  h_{\sigma}{}^{\lambda}{}_{,\beta}{}^{\sigma} - h_{\sigma
    \beta,}{}^{\lambda \sigma} + h_{,}{}^{\lambda}{}_{\beta} \right), 
\nonumber \\
C^{\mu \nu}_D &=& \frac{\dot{f}}{4} \tilde{\epsilon}^{\sigma \alpha \beta
  (\mu} \left( 2 h^{0}{}_{(\sigma,\tau)} - h_{\sigma \tau,}{}^{0} \right)
\left( h^{\tau}{}_{[\beta,\alpha]}{}^{\nu} -
  h^{\nu}{}_{[\beta,\alpha]}{}^{\tau} \right).
\nonumber \\
\ea
The last term of the Cotton tensor is simply the derivative of the
Ricci tensor which we already calculated to second order in
Eq.~(\ref{Ricci-2nd-O}). In order to avoid notation clutter, we shall
not present it again here, but instead we combine all the Cotton
tensor pieces to obtain
\begin{widetext}
\ba
C^{\mu \nu} &=& - \frac{\dot{f}}{2} \tilde{\epsilon}^{0 \alpha \beta
  (\mu} \left( \square_{\eta} h^{\nu)}{}_{\beta,\alpha} - h_{\sigma
    \beta, \alpha}{}^{\sigma \nu)} \right) - \frac{\dot{f}}{2}
\tilde{\epsilon}^{0 \alpha \beta (\mu} \left[ h \left(\square_{\eta}
      h^{\nu)}{}_{\beta,\alpha} - h_{\sigma \beta, \alpha}{}^{\sigma
        \nu)} \right) + \frac{1}{2} \left( 2
      h^{\nu)}{}_{(\lambda,\alpha)} - h_{\lambda \alpha,}{}^{\nu)}
    \right) 
\right. 
\\ \nonumber 
&\times& \left.
\left( \square_{\eta}h^{\lambda}{}_{\beta} - 2
      h_{\sigma}{}^{(\lambda}{}_{,\beta)}{}^{\sigma} +
      h_{,\beta}{}^{\lambda} \right) - 2 \hat{Q}
    R^{\nu)}{}_{\beta,\alpha} \right]  + \frac{\dot{f}}{4}
 \tilde{\epsilon}^{\sigma \alpha \beta (\mu} \left( 2
   h^{0}{}_{(\sigma,\tau)} - h_{\sigma \tau,}{}^{0} \right) \left(
   h^{\tau}{}_{[\beta,\alpha]}{}^{\nu)} -
   h^{\nu)}{}_{[\beta,\alpha]}{}^{\tau} \right) + {\cal{O}}(h)^3
\ea
\end{widetext}
where its covariant form is 
\begin{widetext}
\ba
C_{\mu \nu} &=& - \frac{\dot{f}}{2} \tilde{\epsilon}^{0 \alpha
  \beta}{}_{(\mu} \left( \square_{\eta} h_{\nu) \beta,\alpha} -
  h_{\sigma \beta, \alpha \nu)}{}^{\sigma} \right) - \frac{\dot{f}}{2}
\tilde{\epsilon}^{0 \alpha \beta}{}_{(\mu} \left[ h \left(\square_{\eta}
      h_{\nu) \beta,\alpha} - h_{\sigma \beta, \alpha \nu)}{}^{\sigma}
    \right) + \frac{1}{2} \left( 2 h_{\nu) (\lambda,\alpha)} -
      h_{\lambda \alpha, \nu)} 
    \right) 
\right. 
\nonumber \\
&\times& \left.
\left( \square_{\eta}h^{\lambda}{}_{\beta} - 2
      h_{\sigma}{}^{(\lambda}{}_{,\beta)}{}^{\sigma} +
      h_{,\beta}{}^{\lambda} \right) - 2 \hat{Q}
    R_{\nu) \beta,\alpha} + h_{\nu \lambda} \left(\square_{\eta}
      h^{\lambda)}{}_{\beta,\alpha} - h_{\sigma \beta, 
    \alpha}{}^{\sigma \lambda)} \right)  \right]  
\nonumber \\
&+&
\frac{\dot{f}}{4} \tilde{\epsilon}^{\sigma \alpha \beta}{}_{(\mu}
\left( 2 h^{0}{}_{(\sigma,\tau)} - h_{\sigma \tau,}{}^{0} \right)
\left( h^{\tau}{}_{[\beta,\alpha] \nu)} -
  h_{\nu)[\beta,\alpha]}{}^{\tau} \right) - \frac{\dot{f}}{2} h_{\mu
  \lambda} \tilde{\epsilon}^{0 \alpha \beta (\lambda} \left(
  \square_{\eta} h_{\nu) \beta,\alpha} - h_{\sigma \beta, \alpha
    \nu)}{}^{\sigma} \right) + {\cal{O}}(h)^3. 
\ea
\end{widetext}

For the PPN mapping of CS modified gravity, only the $00$ component of
the metric is needed to second order, which implies we only need
$C_{00}$ to ${\cal{O}}(h)^2$. This component is given by
\ba 
C_{00} &=& \frac{\dot{f}}{4} \tilde{\epsilon}^{i j
  k}{}_{0} \left( 2  h^{0}{}_{(i,\ell)} - h_{i
    \ell,}{}^{0} \right) \left(  h^{\ell}{}_{[k,j] 0} - 
   h_{0 [k,j]}{}^{\ell} \right) 
\nonumber \\
&-& \frac{\dot{f}}{2} h_{0 
   \ell} \tilde{\epsilon}^{0 j k (\ell} \left(
   \square_{\eta} h_{0 k,j} - h_{i  k, j
     0}{}^{i} \right) + {\cal{O}}(h)^3,
\ea
where in fact the last term vanishes due to the PPN gauge condition.
Note that this term is automatically of ${\cal{O}}(6)$, which is well
beyond the required order we need in $h_{00}$.


\end{document}